\documentclass[prl,showpacs,twocolumn,amsmath,amssymb,amsfonts,superscriptaddress,letterpaper]{revtex4-1}

\usepackage{graphicx}
\usepackage{dcolumn}
\usepackage{bm}

\begin{document}

\title{New Determination of Muonium-Deuterium 1$S$-2$S$ Isotope Shift}


\author{Isaac Fan}
\email{ifan@phys.nthu.edu.tw}
\affiliation{Department of Physics, National Tsing Hua University, Hsinchu City, 30013, Taiwan (R.O.C.)}
\affiliation{Frontier Research Center on Fundamental and Applied Sciences of Matters, National Tsing Hua University, Hsinchu City, 30013, Taiwan (R.O.C.)}

\author{Chun-Yu Chang}
\affiliation{Institute of Photonics Technology, National Tsing Hua University, Hsinchu City, 30013, Taiwan (R.O.C.)}

\author{Li-Bang Wang}
\affiliation{Department of Physics, National Tsing Hua University, Hsinchu City, 30013, Taiwan (R.O.C.)}
\affiliation{Frontier Research Center on Fundamental and Applied Sciences of Matters, National Tsing Hua University, Hsinchu City, 30013, Taiwan (R.O.C.)}

\author{Simon L. Cornish}
\affiliation{Joint Quantum Centre (JQC), Durham--Newcastle, Department of Physics, Durham University, Durham, DH1 3LE, United Kingdom}

\author{Jow-Tsong Shy}
\affiliation{Department of Physics, National Tsing Hua University, Hsinchu City, 30013, Taiwan (R.O.C.)}
\affiliation{Frontier Research Center on Fundamental and Applied Sciences of Matters, National Tsing Hua University, Hsinchu City, 30013, Taiwan (R.O.C.)}
\affiliation{Institute of Photonics Technology, National Tsing Hua University, Hsinchu City, 30013, Taiwan (R.O.C.)}

\author{Yi-Wei Liu}
\email{ywliu@phys.nthu.edu.tw}
\affiliation{Department of Physics, National Tsing Hua University, Hsinchu City, 30013, Taiwan (R.O.C.)}
\affiliation{Frontier Research Center on Fundamental and Applied Sciences of Matters, National Tsing Hua University, Hsinchu City, 30013, Taiwan (R.O.C.)}


\begin{abstract}
We report a new determination of muonium 1$S$-2$S$ transition frequency and its isotope shift with deuterium by recalibrating the iodine reference lines using an optical frequency comb.
The reference lines for the muonium and deuterium 1$S$-2$S$ transitions are determined with a precision of 2.4$\times$10$^{-10}$ and 1.7$\times$10$^{-10}$ respectively.
A new muonium-deuterium 1$S$-2$S$ isotope-shift frequency is derived from these references to be 11 203 464.9(9.2)(4.0) MHz, in agreement with an updated bound-state quantum-electrodynamics prediction based on 2010 adjustments of Committee on Data for Science and Technology and 2.3 times better in the systematic uncertainty than the previous best determination.
\end{abstract}
\pacs{to be determined}
\maketitle

The deviation of the proton charge radius between the muonic hydrogen ($\mu$H) Lamb shift measurement  \cite{Pohl2010-nature} and the recommended values in CODATA  \cite{Mohr2006-rmp,Mohr2012-jpcrd} has recently been re-enforced by the 2$S_{1/2}$-2$P_{3/2}$ measurement in the same $\mu$H system  \cite{Antognini2013-science}.
One possible theoretical explanation for the size puzzle is unknown quantum-electrodynamics (QED) corrections on the order of 310 $\mu$eV causing $\mu$H results to be wrongly attributed to the nuclear size effect  \cite{Carlson2012-prd}.
Other postulates for the new interactions have also been suggested  \cite{Barger2012-prl,Batell2011-prl}.

Experimentally, there is another long-standing 3.3$\sigma$ experiment-theory discrepancy in the muon anomalous magnetic moment $(g-2)_{\mu}$ \cite{Brown2001-prl,Bennett2004-prl,Hagiwara2011-jpg,Jungmann2012-hi}.
In addition, the isotope-shift measurement of 1$S$-2$S$ transition between muonium and deuterium is only marginally in agreement with the current theory with a 1.4 ppm deviation.
It is tentative, therefore, to speculate that new interactions may be of muon-related origin.

Microwave and laser spectroscopy of the muonium atom, a purely two-body leptonic bound-state (Mu, $\mu^+e^-$),  can offer stringent experimental tests for the bound-state QED \textit{without the finite-size effect} due to the structureless muon nucleus \cite{Hughes1990-book,Liu1999-prl,Meyer2000-prl}.
This removes the main limiting factor caused by the hadronic structure in $\mu$H or H when comparing the theory and the experiment.

Among the lower lying levels of non-Rydberg state Mu atoms, the electromagnetic 1$S_{1/2}$-2$S_{1/2}$ transition is of particular importance because the fundamental property of a muon (e.g. mass) can be inferred from it \cite{Meyer2000-prl}.
Currently, the most accurate value of the muon mass is $m_{\mu}/m_{e}$=206.768 2843(52) suggested by CODATA \cite{Mohr2012-jpcrd}, which derived its small uncertainty from the Mu ground-state (1$S$) hyperfine splitting \cite{Liu1999-prl}.
The natural linewidths of the Mu 1$S$ hyperfine and the 1$S$-2$S$ transitions are both 145 kHz, limited by the $\approx$ 2.2 $\mu s$ lifetime of the muon. 
Therefore, the optical (higher frequency) 1$S$-2$S$ transition should in principle offer orders of magnitude higher accuracy than the microwave (lower frequency) ground-state hyperfine transition.

In the last 1$S$-2$S$ isotope-shift measurement done at ISIS muon facility of the Rutherford Appleton Laboratory in UK \cite{Meyer2000-prl}, however, this optical advantage was not obvious because the measurement was statistically limited by the low vacuum yield of the muonium source ($\Delta f$= 9.2 MHz) and systematically limited by the low accuracy of the deuterium reference line ($\Delta f$= 9.3 MHz).

In this Letter, we carry out a frequency comb calibration of the iodine reference cell used in the ISIS experiment with a Doppler-free saturation spectrometer.
Combining our calibration with the experimental parameters of the ISIS measurement, we can reduce the systematic uncertainty by 2.3 times.
The calibration leads to an updated Mu-D 1$S$-2$S$ isotope-shift of 11 203 464.9(9.2)(4.0) MHz, where the first bracket indicates the statistical uncertainty and the second indicates the systematic uncertainty. 
This is in better agreement with the current theory.

\begin{figure}[t]
	\includegraphics[width=1\linewidth]{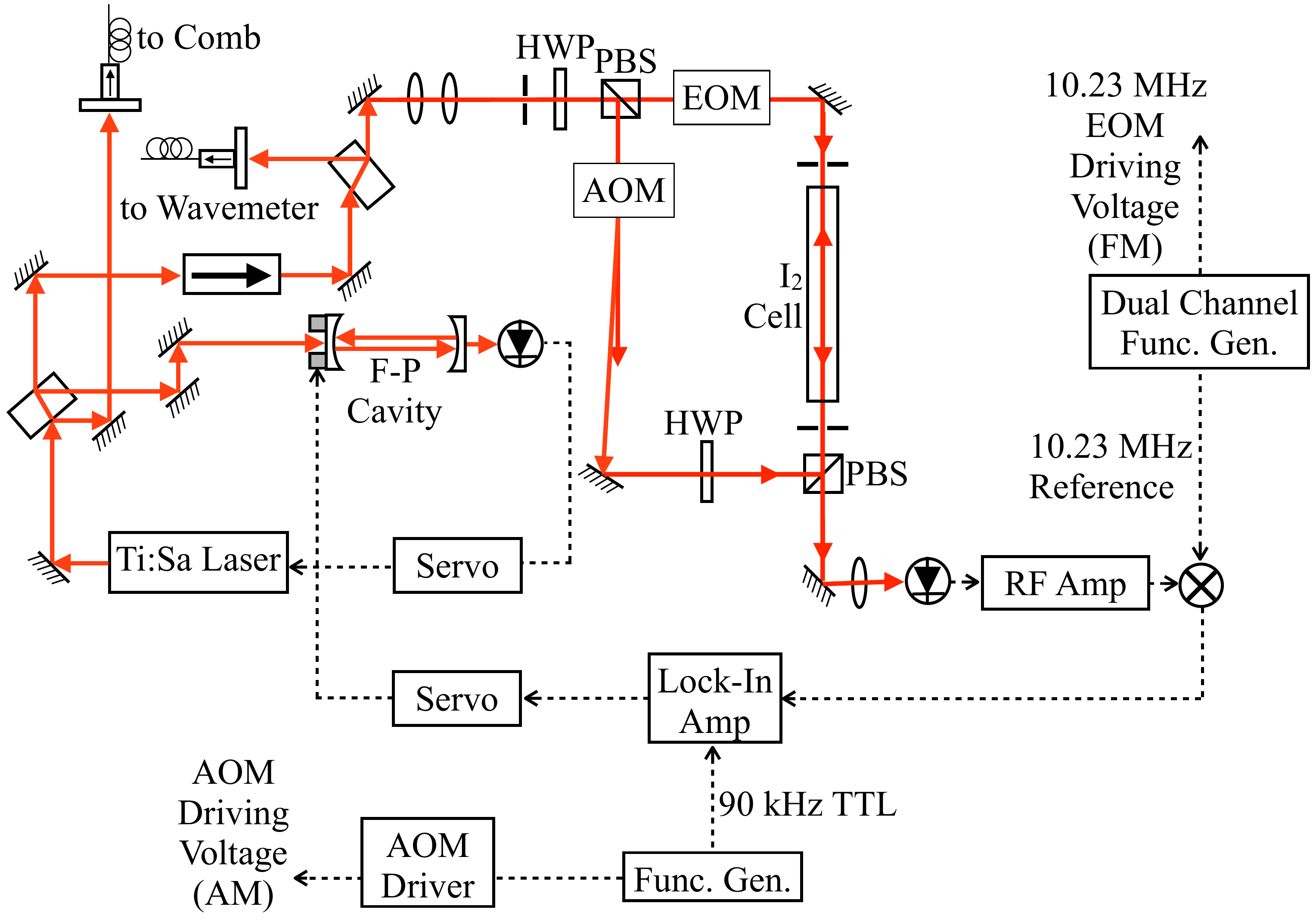}
	\caption{\label{fig:layout}(Color Online) The schematic of this experiment. HWP: half waveplate; PBS: polarizing beam splitter; F-P cavity: Fabry-Perot cavity; AOM: acoustic-optical modulator; EOM: electro-optical modulator.}
\end{figure}

Our experimental setup is shown in Fig.~\ref{fig:layout}.
The 730 nm light source ($\approx$ 500 mW) was from a titanium-sapphire laser (Technoscan TIS-SF-07e) pumped by a 10 W diode-pumped solid-state laser (Coherent Verdi V10). 
The light beam was subsquently diverted into (a) the Fabry-Perot cavity (FSR$\approx$ 1 GHz) for the frequency stabilization and scanning, (b) the wavemeter (resolution$\approx$ 1 GHz) for monitoring, (c) the optical freqency comb (OFC) for absolute frequency calibration, and (d) the cell area for resolving the Doppler-free spectral features which was partially based on our previous frequency-modulation saturation spectrometer \cite{Cornish2000-josab}.

The beam delivered to the cell area was collimated ($\approx$ 1 mm in radius) and further branched into the pump ($\approx$ 185 mW or equivalently $\approx$ 5.89$\times 10^4$ W/m$^2$) and the probe ($\approx$ 2 mW) whose powers were adjusted independently.
The pump beam was amplitude modulated (AM) at 90 kHz with an acoustic-optical modulator (IntraAction ATM-801A2).
The probe beam was phase modulated (PM) with an resonant-type electro-optical modulator at 10.23 MHz (LC resonant circuit plus a Thorlabs EO-PM-NR-C2) before passing through the cell and being detected by a fast photoreceiver (125 MHz New Focus 1801).
The phase modulation depth was set such that the carrier amplitude is $\approx$ 6 times bigger than the amplitudes of the 10.23 MHz sidebands.
These amplitude and phase modulation parameters are identical to the conditions used in the ISIS measurement \cite{Meyer2000-prl,Cornish2000-josab}.

The mixer (Mini-Circuits ZAD-1-1) demodulated the preamplified (HP 8447D) signal which then was subsequently processed by the lock-in amplifier (Stanford Research SR830) to demodulate the 90 kHz signal with an adjustable phase, sensitivity, and time constant.
The laser frequency was servoed (10 MHz Precision Photonics LB1005) to the center of the demodulated hyperfine signal and was calibrated by fiber coupling some laser light to the optical frequency comb (OFC).
The usage of our OFC has been described previously \cite{Fan2011-pra}.
The accuracy of our OFC system was verified by measuring the a$_{10}$ line of the $^{127}$I$_2$ molecular iodine R(56)32-0 transition ($\approx$ 532 nm) with an error limit less than 200 kHz \cite{Liao2010-josab}. 

\begin{figure}[t]
	\includegraphics[width=0.8\linewidth]{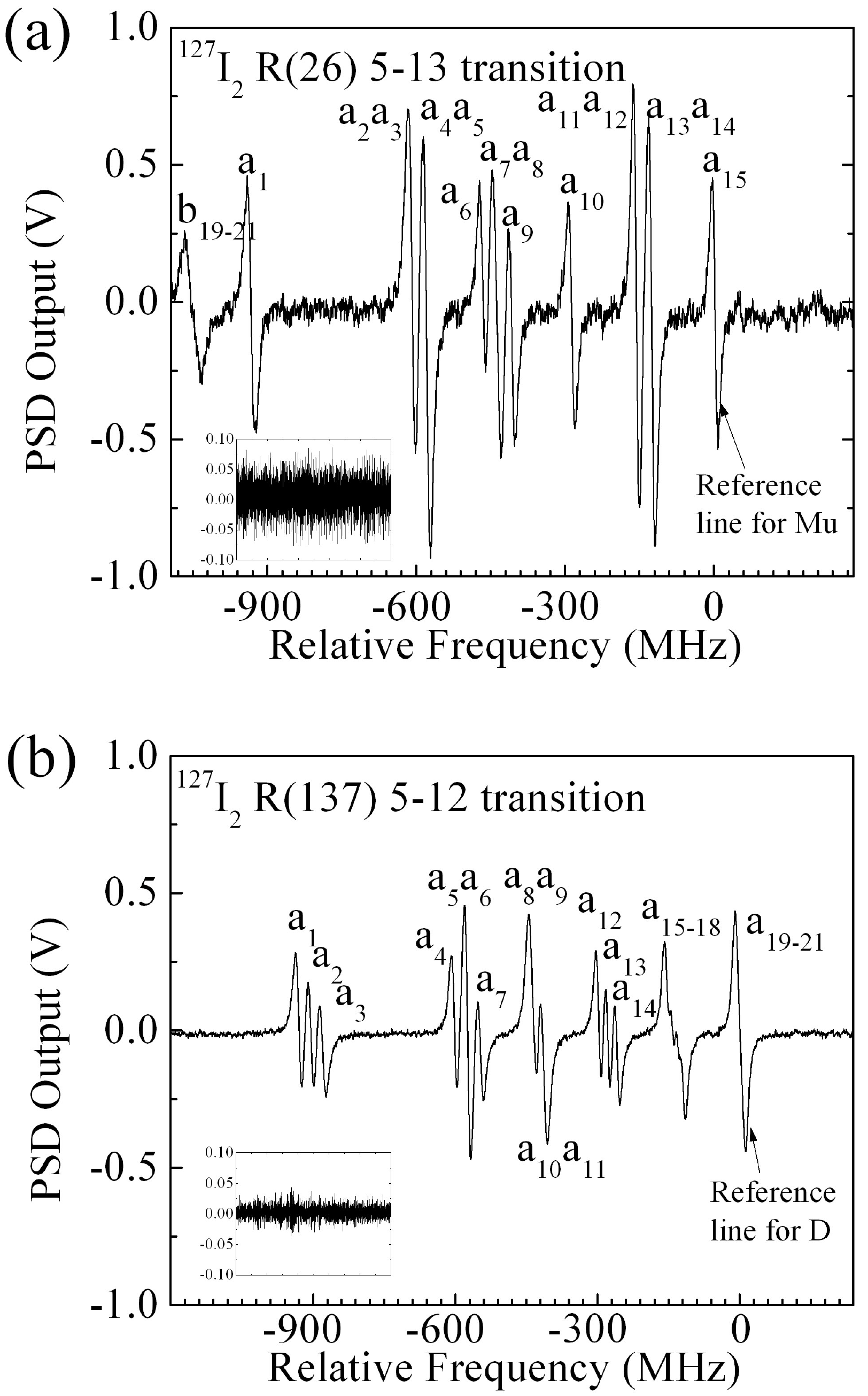}
	\caption{\label{fig:spectra}(a) and (b) are the hyperfine spectra for Mu and D respectively (along with frequency markers) obtained at an oven temperature of 650 $^{\circ}$C following Ref.~\cite{Meyer2000-prl}. Figure insets are the stabilized phase-sensitive detection (PSD) outputs for a 10 s duration. The cold finger temperatures of the cell are set at 43 $^{\circ}$C and 33 $^{\circ}$C for Mu and D respectively. The experimental linewidth is 11.2 MHz for Mu and 23.0 MHz for D. The raw signal-to-noise ratio (SNR) is 41.2 for Mu and 98.3 for D. A time constant of 10 ms and 6dB/oct slope is used in both spectra which lead to an equivalent noise bandwidth (ENBW) of 25 Hz. Therefore, the calibrated 1 Hz noise bandwidth SNR are 206 $\sqrt{Hz}$ and 491 $\sqrt{Hz}$ for Mu and D respectively.}
\end{figure}

\begin{figure}[b]
	\includegraphics[width=0.9\linewidth]{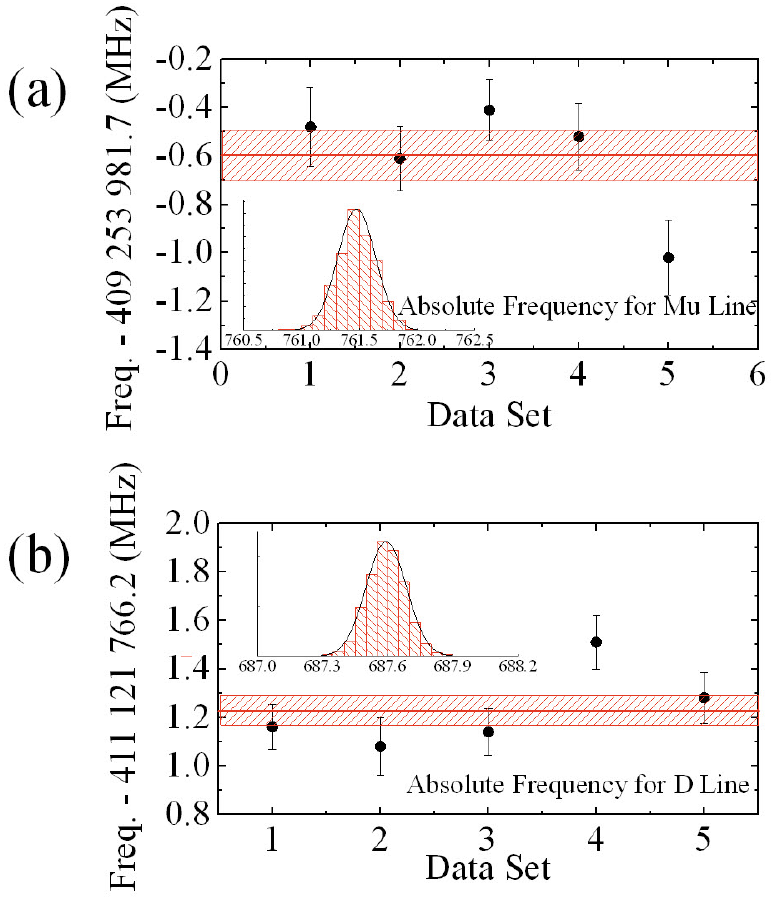}
	\caption{\label{fig:combresults}(Color Online) Absolute frequency comb measurements for (a) Mu and (b) D respectively. The red lines represent constants from the zero-slope linear fitting of each data set. The shaded areas indicate the standard error of fits. The fitted values are Mu: -0.6(0.1) MHz and D: 1.23(07) MHz. The insets are typical histograms of the beat frequencies in MHz.}
\end{figure}

\begin{table*}[bth]
	\caption{\label{tab:summary}Summary of QED theoretical predictions on the H, Mu, and D 1$S$-2$S$ frequencies and their comparison to experimental values. The CODATA-10 adjustments of basic constants are used. Only the dominating uncertainty contributions from the electron and nucleon masses and sizes are included in the calculation. Wherever applicable, the first uncertainty bracket indicates the statistical uncertainty and the second bracket indicates the systematic uncertainty.}
	\centering
	\begin{tabular}{c | c | c | c | c}
	\hline\hline
					&1$S$-2$S$ Contributions & Hydrogen (MHz) 		& Muonium (MHz)          	& Deuterium (MHz)\\
	\hline\hline	
	Theory			&Dirac Eigenvalue  		&2 466 068 541.005 71(75)  	&2 455 535 991.79(40) 	&2 466 739 545.088 35(37) \\
	(CODATA-10)				&Lamb Shift           		&-7 126.786 097(13)      	&-7 056.046 727(17)        &-7 131.300 102(81)\\
					&Finite Size Effect     	&-1.054(14)          		&0          				&-6.288(14)\\
					&&&&\\
					&&&&\\
					&Total 	&2 466 061 413.165(15)  	&2 455 528 935.74(40)  	&2 466 732 407.500(15)\\
					&&&&\\
	\hline
	Experiment		&this work   &                     	&2 455 528 940.6(9.1)(3.7)&2 466 732 405.5(1.1)(1.5)\\
	                  		&Ref.~\cite{Meyer2000-prl}   	&                             	&2 455 528 941.0(9.1)(3.7)&2 466 732 397.2(1.1)(8.5)\\
	                			&Ref.~\cite{Parthey2011-prl} 	&2 466 061 413.187 035(10) 	&&\\
					&Ref.~\cite{Parthey2011-prl}+Ref.~\cite{Parthey2010-prl} 	&				&		&2 466 732 407.521 641(18)\\
	\hline\hline
	\end{tabular}
\end{table*}

The 54.1 cm long iodine absorption cell used for the ISIS measurement was manufactured by the University of Heidelberg (code named HEID4 in Ref.~\cite{Cornish1998-thesis}) in the 1990's which we retrieved recently for this comb calibration project.
The cell was made of quartz that can withstand temperatures up to 700 $^{\circ}$C.
The heating of iodine vapor to above 500 $^{\circ}$C was necessary in order to access the rovibrational transitions of the \textit{B-X} system originating from high-lying vibrational levels of the electronic ground state \cite{Cornish2000-josab,Gerstenkorn1980-jp,footnote1}.
Our oven temperature dependence study showed that while the population of the high-lying levels of the ground state depended on the oven temperature (evidenced by a change in the signal-to-noise strength at different oven temperatures), the frequencies of these hyperfine transitions remained unchanged.

Two weak molecular $^{127}$I$_2$ transitions, not reported in the iodine atlas at 13 350-13 920 cm$^{-1}$ \cite{Gerstenkorn1982-atlas}, were of interest corresponding to the Mu and D 1$S$-2$S$ transitions.
One important criterion for these reference lines was that they should be located \textit{within 1 GHz of the target transition} to allow the use of acoustic-optical modulators (AOM) to bridge the frequency offset. 
For the Mu 1$S$-2$S$ reference, the a$_{15}$ line of the P(26) 5-13 transition was calibrated.
For the D 1$S$-2$S$ reference, the a$_{19-21}$ line of the P(137) 5-12 transition was calibrated.
Figure \ref{fig:spectra} shows both reference lines. 

An estimate of the frequency stability $\Delta f$ can be obtained by dividing the linewidth by the SNR, leading to $\Delta f_{Mu}\approx$ 270 kHz and $\Delta f_D\approx$ 230 kHz.
The non-zero means of the Gaussian fits to the frequency stabilized PSD outputs (insets of Fig.~\ref{fig:spectra}) result in an offset frequency of -70 kHz for both Mu and D. 
The lineshape asymmetries of the hyperfine signals contribute to an additional offset -450 kHz and -80 kHz.
The net systematic offset is thus -520 kHz and -150 kHz for Mu and D respectively indicating the distance of measured frequencies away from the true hyperfine centers.

Figure~\ref{fig:combresults} shows the absolute frequency calibrations of the two isotopic 1$S$-2$S$ reference lines.
Each data point shown in Fig.~\ref{fig:combresults} is a normal distribution-fitted mean of 1000 beat frequencies (gated 0.1 s, $\geq$30 dB in strength) with the error bar indicating 1$\sigma$ (one standard deviation) of the distribution.
The standard deviation is higher for the Mu reference line in agreement with the relative SNR sizes of the observed hyperfine spectra (Fig.~\ref{fig:spectra}).

The absolute frequency is calculated using the following equation:
\begin{eqnarray}
 f_{\rm{measure}}=N\times f_{\rm{rep}}\pm f_{\rm{off}}\pm f_{\rm{beat}}-f_{\rm{AOM}}/2.\nonumber 
\end{eqnarray}
where $f_{\rm{rep}}$, $f_{\rm{off}}$, and $f_{\rm{beat}}$ are the repetition rate, the offset frequency, and the beat frequency of the OFC.
$f_{\rm{AOM}}$ is the frequency shift of the AOM incurred during the amplitude modulation of the pump beam.
The re-calibrated reference values are (taking systematic offsets into account):
\begin{eqnarray}
 \mbox{a$_{15}$, P(26) 5-13 (Mu)}&=&\mbox{409 253 981.6(0.1) MHz},\nonumber \\
 \mbox{a$_{19-21}$, P(137) 5-12 (D)}&=&\mbox{411 121 767.58(07) MHz}.\nonumber
\end{eqnarray}
From these recalibrated values, the 1$S$-2$S$ energy intervals are updated in Table~\ref{tab:summary} along with theoretical values.
Since the 1$S$-2$S$ spectroscopy in Mu and D involves a frequency tripling from the baseband followed by a two-photon absorption, the uncertainty in the 1$S$-2$S$ reference lines would be 6 times that of the iodine reference lines \cite{Meyer2000-prl}. 

Among the various systematic contributions of the ISIS Mu 1$S$-2$S$ frequency measurement (Table I of Ref.~\cite{Meyer2000-prl}), the residual linear Doppler shift and line-fitting contributed 3.4 MHz and 1.2 MHz respectively.
Although our OFC calibration improves the Mu 1$S$-2$S$ standard to 0.6 MHz (6$\times$0.1 MHz), it is too small of a correction to influence the final systematic uncertainty significantly.
This is the reason that our updated Mu 1$S$-2$S$ frequency remains essentially the same as the ISIS measurement.

On the other hand, our OFC calibration has improved the uncertainty of the D 1$S$-2$S$ reference standard from 8.4 MHz to 0.42 MHz, a 20 times improvement.
This systematic reduction improves the final uncertainty significantly.
The final systematic uncertainty of D 1$S$-2$S$ frequency is updated to 1.5 MHz.
In addition, the difference with the most accurately determined D 1$S$-2$S$ experimental value (extracted from the continuous-wave spectroscopy in Ref.~\cite{Parthey2011-prl} and Ref.~\cite{Parthey2010-prl}) is reduced to 0.82 ppb in comparison to 4.2 ppb of the ISIS measurement.

\begin{table}[t]
	\caption{\label{tab:is}List of Mu-D 1$S$-2$S$ isotope-shift intervals.}
	\centering
	\begin{tabular}{c | c }
	\hline\hline
		& 1$S$-2$S$ Mu-D Isotope-shift Interval (MHz) \\
		\hline
		theory & 11 203 471.8(0.4)  \\
		Ref.~\cite{Meyer2000-prl}(exp.) & 11 203 456.2(9.2)(9.3)  \\
		this work & 11 203 464.9(9.2)(4.0) \\
	\hline\hline
	\end{tabular}
\end{table}

Taking the difference between the Mu 1$S$-2$S$ and D 1$S$-2$S$ frequencies, the theoretical and experimental Mu-D 1$S$-2$S$ isotope-shift frequencies are summarized in Table~\ref{tab:is}.
It is worth noting that the D 1$S$-2$S$ frequency interval from the continuous-wave spectroscopy measurement is not suitable for this purpose since it did not experience the same systematics in the same apparatus as the Mu 1$S$-2$S$ measurement, which is important for the isotope-shift determination.
The 400 kHz uncertainty of the theoretical Mu-D isotope-shift mainly comes from the electron-muon mass ratio measurement \cite{Liu1999-prl}.
The good agreement between our updated Mu-D isotope-shift interval with the theory indicates that the systematic effect in the ISIS measurement has been correctly taken into account.
The systematic uncertainty is improved by 2.3 times and the difference with the theoretical value is changed from 1.4 ppm to 0.62 ppm.
This is the most accurate experimental determination of Mu-D 1$S$-2$S$ isotope-shift frequency.

In summary, we have calibrated the $^{127}$I$_2$ reference cell, used in the last Mu-D 1$S$-2$S$ isotope-shift measurement, using a frequency comb. 
We determine a new value for the Mu-D 1$S$-2$S$ isotope-shift frequency which agrees with the updated value calculated with CODATA fundamental constants.
At the present level of experimental accuracy, no inconsistency with the current bound-state QED theory can be found.

Further improvement in the experimental accuracy will require increasing the statistics.
Recent technological advancement in increasing the Mu vacuum yield to 38\% at 250 K with mesoporous silicon \cite{Antognini2012-prl}, which may permit continuous-wave (cw) laser spectroscopy in an enhancement cavity, shows great promise in alleviating this statistical limitation.

We would like to thank the support from the National Science Council of Taiwan and the National Science Council of Taiwan-Royal Society of UK International Exchange Program (NSC 99-2911-I-007-027).

%

\end{document}